\documentclass[11pt,twoside]{article}

\usepackage{asp2006}
\usepackage{epsf}
\usepackage{lscape}

\begin{document}
\title{Effects of Jet Opening Angle and Velocity Structure on Blazar Parameters}   %%% Fill in title
\author{Paul J. Wiita,\altaffilmark{1} Gopal-Krishna,\altaffilmark{2}
Samir Dhurde,\altaffilmark{3} and Pronoy Sircar\altaffilmark{4}}   %%% Fill in author names
%\affil{}    %%% Fill in author affiliations
\altaffiltext{1}{Department of Physics \& Astronomy, Georgia State University,
P.O.\ Box 4106, Atlanta, GA 30302-4106, USA}
\altaffiltext{2}{National Centre for Radio Astrophysics/TIFR, Post Bag 3,
Pune 411007, India}
\altaffiltext{3}{Inter-University Centre for Astronomy \& Astrophysics,
Post Bag 4,  Pune 411007, India}
\altaffiltext{4}{Department of Physics, Indian Institute of Technology, Kanpur 208016, India} 
\begin{abstract} %%% Abstract to run on from here.
We had earlier shown that for a constant velocity jet the discrepancy between the low speeds ($\beta$) indicated by VLBI knot motions and the  high Doppler factors 
($\delta$) inferred from
emission of TeV photons could be reconciled if ultrarelativistic jets possessed
modest opening angles.  Here we evaluate the 
(flux-weighted) viewing angles of the jet and the apparent $\beta$ and 
$\delta$ values of the radio
knots on parsec scales.   The influence
of the jet opening angle on these radio knot parameters are found 
for the usually considered types of relativistic nuclear jets: those with
uniform bulk speeds and those where the bulk Lorentz factor of the flow 
decreases with distance from the jet axis, known as ``spine--sheath'' flows.  For both
types of jet velocity structures the expectation value of the jet orientation
angle at first falls dramatically with increases in the (central) jet Lorentz 
factor, but for extremely relativistic jets it levels off at a fraction of the opening angle.  The effective values of the apparent speeds and Doppler 
factors of the knots always decline substantially with increasing jet opening
angle.  The rarity of highly superluminal parsec-scale radio components in TeV blazars can be understood if their jets are both highly relativistic and intrinsically weaker, so probably less well collimated, than the jets in ordinary blazars.
\end{abstract}
%%% MAIN BODY OF TEXT GOES HERE. CONSULT "INSTRUCTIONS FOR AUTHORS USING
%%% LATEX2E MARKUP", SECTIONS 2.3-2.6 FOR HELP WITH EQUATIONS, FIGURES,
%%% AND TABLES.
%\section{}   %%% Top level section head (remove "%" symbol)
%\subsection{}   %%% Second level section head (remove "%" symbol)
%\subsubsection{}   %%% Lowest level section head (remove "%" symbol)
%%% CONSULT SECTION 3 OF "INSTRUCTIONS FOR AUTHORS" FOR HOW TO USE NATBIB.
%%% AUTHORS ARE ENCOURAGED TO USE EITHER THE "THEBIBLIOGRAPY" ENVIRONMENT
%%% BY UNCOMMENTING (DELETING THE "%" SYMBOL) THE COMMANDS BELOW, OR BY
%%% USING THE BIBTEX ENVIRONMENT. TO FIND OUT WHICH IS APPLICABLE TO YOUR
                                                                                   
\section{Introduction}
\label{sec:intro}
While
Very Long Baseline Interferometry (VLBI) monitoring of the radio knots in blazar jets has found
several sources containing knots with apparent speeds $v_{app}$ in excess of 25$c$,
the typical speeds for blazars known to emit the highest energy $\gamma$-ray photons 
(TeV blazars) are found to be much more modest, with $v_{app} < 5c$, 
and their radio knots are actually often subluminal \citep[e.g.,][]{pine04}.
But very large bulk Lorentz factors in these 
parsec-scale blazar jets ($\Gamma > 30$) are inferred from TeV flux 
variations \citep[e.g.,][]{kraw02}.
Ultra-relativistic 
bulk Lorentz factors have also been inferred from the intraday radio 
variability of some blazars \citep[e.g.,][]{macq06} and from VLBI measurements of the brightness 
temperatures of several blazar nuclei \citep[e.g.,][]{hori04}.
This large difference between the estimates of $\Gamma$ 
derived from these observations led some authors 
to postulate a dramatic jet deceleration between sub-parsec and parsec 
scales \citep[e.g.,][]{geor03}.
Another approach to this contradiction has been to invoke a ``spine--sheath''
configuration for the jets; the fast spine close to the jet axis yields
the $\gamma$-ray emission and the surrounding slower moving 
sheath produces the radio emission  
\citep[e.g.,][]{ghis05}.

We recently undertook an analytical study
which showed that the modest apparent speeds of the knots of blazars, 
which are mostly unresolved by VLBI, 
can be reconciled with the extremely relativistic bulk motion 
inferred for TeV, and some other, blazars, if one 
considers a modest full opening angle $(\omega \sim 5^{\circ}$--$10^{\circ})$ 
for the parsec-scale jets \citep[][hereafter, Paper1]{gopa04}.  
We also showed that the actual viewing angles, $\theta$, of such 
{\it conical} jets from the line-of-sight can be substantially larger than 
those commonly inferred \citep[e.g.,][]{jors05} by combining the flux variability and the VLBI proper 
motion data \citep[][Paper II]{gopa06}. Direct support 
for the assumption of conical parsec-scale jets comes from the VLBI 
imaging of the nuclear jets in the nearest two radio galaxies, M87 
 and Centaurus A.  
 
Here we  summarize  the predictions for 
two widely discussed jet forms (i.e., those with uniform $\Gamma$ and 
those with velocities decreasing away from the jet axis, the spine--sheath 
types) for how three key parameters of the VLBI radio knots 
in blazar jets, the apparent speed ($\beta_{app}$), Doppler factor ($\delta$), and viewing angle $(\theta)$, would be influenced by the jet opening angle.  The details of this work
have recently appeared in \citet{gopa07a}, and kinematical
diagrams of $\delta$ against $\beta$ for conical jets are in \citet{gopa07b}.  

\section{Finding  Apparent Parameters for Conical Jets}
\label{sec:uniform and spine--sheath jets}

We approximate a 
radio knot with a circular disk-shaped region of uniform intrinsic 
synchrotron emissivity, and thus uniform surface brightness, presumably arising from a shock in the jet. 
We define the angle between  the axis of the jet
and the line-of-sight from the stationary core of the blazar to be
$\theta$ and the velocity of the knot along the jet to be $\vec{\beta} \equiv \vec{v}/c$.  Jets with finite opening angles require  more general forms of the usual relations: 
$\delta = [\Gamma (1-\vec{\beta} \cdot \hat{n})]^{-1}$, 
where  $\Gamma= (1-\beta^2)^{-1/2}$, $\hat{n}$ is the unit vector along the sight line, and
$\vec{v}_{app} = \vec{v} \times \hat{n}/(1-\vec{\beta} \cdot \hat{n}).$
The values
of $\delta$ and $v_{app}$ arising from each $d\Omega'$ patch vary when $\omega \neq 0$.  We generalize those results by noting that the magnitude
of the bulk velocity,
$c\beta$,  might also vary as a function of the angular distance ($r$) of that patch from the jet axis.  

The {\it effective} values, $\beta_{\rm eff}$ and 
$\delta_{\rm eff}$, for a knot, as would be determined from VLBI observations unable to resolve the width of the jet (i.e., knot) are: 
\begin{equation}
S_{obs} = \int_{\Omega} \delta^{p}(\Omega')S_{em}(\Omega')d\Omega' \equiv
A(\theta)S_{em}; 
\end{equation}
\begin{equation}
\delta_{\rm eff} = A^{\frac{1}{p}}(\theta);~~~ 
%\end{equation}
%\begin{equation}
\vec{\beta}_{\rm app,eff} = \frac{1}{S_{obs}} \int_{\Omega} \vec{\beta}(\Omega^\prime)\delta^{p}
(\Omega^\prime)S_{em}(\Omega') d\Omega'.
\end{equation}  
Here, $S_{em}$ and $S_{obs}$ are the emitted and the (Doppler boosted) 
observed flux densities, respectively, $\Omega^{\prime} = \vec{\beta}\cdot{\hat{n}}/\beta$ denotes the location on the knot given by the direction cosine between the local velocity and the line of sight to the nucleus, $\Omega$ is the entire solid angle subtended by the  knot, $A(\theta)$ is the flux boosting factor averaged 
over the  radio knot's  cross-section, and $p$ is usually taken as 3, appropriate for a compact discrete source with flat spectral index. 
The viewing angle, $\theta$, to the jet axis 
measures the angular offset of the circular radio disk's center
from the 
direction of the AGN core. The probability distribution of the viewing
angle, ${\cal P}(\theta)$, is used to compute the expectation value of 
$\theta$ for any particular combination of $\Gamma, \omega, p$ and $q$, with 
the last parameter defined below.

\begin{figure}
\plotone{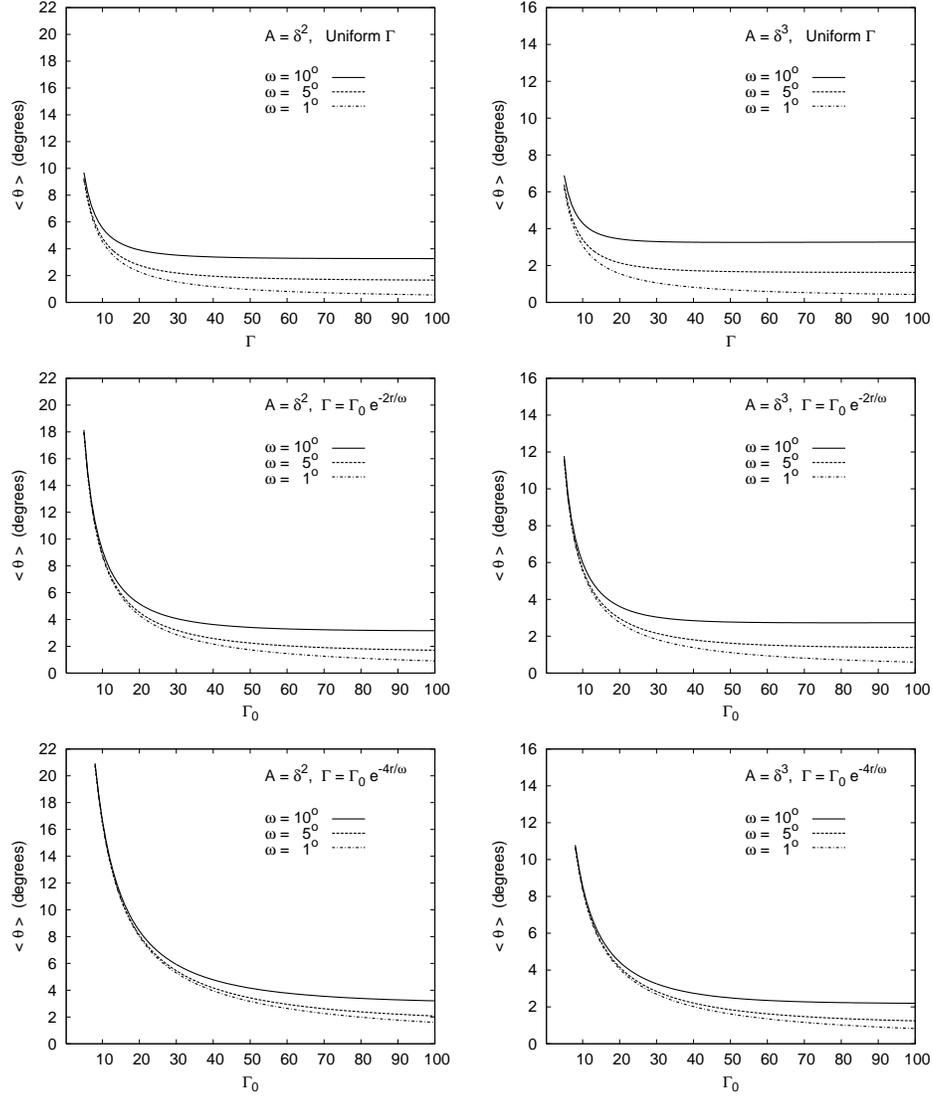}
\caption{Expectation values of the viewing angle, $\left<\theta\right>$, against the jet (central) Lorentz factor, $\Gamma_0$, for uniform (top panel)
and transversely structured (middle and lower panels) jets. The left panels 
correspond to $p=2$ and the right to $p=3$. Solid, dashed and dot-dashed lines are for 
$\omega = 10^\circ, 5^\circ$, and $1^\circ$, respectively.  Reproduced by permission from \citet{gopa07a}; \copyright Royal Astronomical Society.}
\end{figure}
  
The  shock is taken 
to cover the jet's full cross-section with uniform intrinsic emissivity 
for all cases.  As there is no well supported
form for the gradient  in the literature, we adopt an exponential approximation, 
$\Gamma (r) = \Gamma_0 e^{-2rq/\omega}$,
where $r$ is the angular separation from the centre of the radio knot.
We evaluated results for such stratified jets, taking two values for $q$ (1 and 2). The 
 minimum $\Gamma_0$ allowed is such
that the corresponding $\Gamma (r = \omega/2) > 1$.

The
probability of finding a jet at a viewing angle $\theta$, in a
flux-limited sample, is given approximately by \citep[Paper I;][]{cohe89}:
${\cal P}(\theta)d\theta \propto \sin \theta~
A^{3/2}_{\rm eff}(\theta)~d\theta$,
where the exponent $3/2$ is the typical slope of integral source
counts at centimetre wavelengths 
and $A_{\rm eff}$ was taken from the present computations (Eq.\ 1).  This
was numerically evaluated from $\theta = 0^\circ$ to $90^\circ$ and then used to
compute the expectation values $\left<\theta\right>$ for different combinations of
$\Gamma_0$, $\omega$, $p$ and $q$.
The numerical computations were made 
for both uniform $\Gamma$ jets ($q = 0$) and for the stratified jets ($q = 1$ and $2$). 

\section{Discussion}

Fig.\ 1 shows a steep initial decline of $\left< \theta \right>$ with 
increasing $\Gamma_0$, for all values 
of $\omega$ or $q$.  This decline continues until the regime of ultra-relativistic 
jets ($\Gamma_0 > 30$) is reached, when $\left<\theta\right>$ 
becomes essentially independent of $\Gamma_0$.  The near constancy of $\left< \theta \right>$ for extremely 
relativistic jets is particularly striking for the conical jets having a 
uniform $\Gamma$ and at least a moderate opening angle 
($\omega \sim 5^{\circ}$--$10^{\circ})$.  
It is clear that in a typical radio flux-limited sample of 
blazars, a larger opening angle of an ultra-relativistic jet would 
correspond to a considerably larger $\left< \theta \right>$; hence a milder 
foreshortening due to projection, compared to that obtained from the assumption of a pencil jet ($\omega = 0$), is typically expected (see Paper II).

We now discuss the influence of $\omega$ on the {\it effective} apparent 
speed of a radio knot,  c$\beta_{\rm eff}$, and its {\it effective} 
Doppler factor, $\delta_{\rm eff}$ (Eqs.\ 2).  
The decline of $\beta_{\rm eff}$ with $\omega$ 
is sharper for the knots associated with jets of higher $\Gamma$ (for both 
uniform and stratified types). On the other hand, for well collimated jets 
(i.e., $\omega < 0.5^{\circ}$), $\beta_{\rm eff}$ for a uniform $\Gamma$ 
jet would typically be 1.5 to 2 times higher than if $q=1$, and
between 2 and 4 times higher than if $q=2$, 
cases.  Thus, the fastest {\it spine} component of the jet flow, which 
is near the jet axis, would be substantially concealed in the VLBI measurements. 

Note that the much sharper fall of $\beta_{\rm eff}$ with 
$\omega$ found for the uniform $\Gamma$ case insures that 
already for modest jet opening angles, the value for $\beta_{\rm eff}$ of such 
jets would drop below the corresponding values for both of the spine--sheath 
models. 
While for well
collimated ultra-relativistic jets $\beta_{\rm eff}$ is more strongly suppressed
(relative to uniform $\Gamma$) when $\Gamma$ is more peaked towards the
axis (i.e., $q =2$), the opposite is found for the jets having a 
significant opening angle ($\omega \sim 2^{\circ}$ to $\sim 10^{\circ})$, with the exact 
cross-over value of $\omega$ depending 
on $\Gamma_0$ and $p$. 
In summary, if the conical jet is moderately wide 
($\omega \simeq 10^{\circ}$),  the measured apparent speeds of the VLBI 
knots typically remain under $10c$ for stratified jets as
well as for uniform velocity jets, even if $\Gamma_0$ were extremely 
large ($\simeq 100$). 

Consider the $\omega$-dependence of the 
$\delta_{\rm eff}$ for the different transverse velocity distributions.
We find that for cases of 
very good collimation ($\omega \le 0.5^{\circ}$), the uniform $\Gamma$ jets 
would typically have 2 to 4 times larger $\delta_{\rm eff}$ compared to the 
stratified jets, implying roughly an order-of-magnitude stronger Doppler boost. 
As expected for stratified jets, a sharper spine--sheath contrast (i.e., a larger
$q$) leads to 
a lower $\delta_{\rm eff}$. 
A significant reduction 
in $\delta_{\rm eff}$ with $\omega$ is the typical expectation for both kinds of jet, 
the dependence being stronger for extremely relativistic jets, particularly 
the uniform $\Gamma$ type. 
Extremely relativistic jets of both uniform and stratified $\Gamma$ types 
end up with comparable $\delta_{\rm eff}$ values, once $\omega$ exceeds about 
$5^{\circ}$.
These still very high values of $\delta_{\rm eff}$ mean that rapid
variability in TeV $\gamma$-ray emission is to be expected in any of our ultrarelativistic models as the
variability timescale is proportional to $\delta_{\rm eff}^{-1}$.  

But why is the Lorentz factor dichotomy so striking only 
for TeV blazars \citep[e.g.,][]{pine04}?  All blazars 
show two correlated peaks in their spectral energy distributions and are 
now usually classified by the frequency at which the lower frequency 
(synchrotron) peak occurs.
According to the popular scheme unifying  high 
energy peaked (HBL) and low energy peaked blazars (LBL) 
\citep[e.g.,][]{foss98}, TeV emission, which is an HBL characteristic, would 
be more common among lower luminosity blazars.  

On kpc-scales 
extragalactic jets tend to be less well collimated for lower luminosity sources.
If this correlation holds on sub-pc scales, 
then the connection between HBL properties and intrinsically weaker jets would mean that the jet opening angle should 
be larger for TeV emitting jets.  This implies that 
the probability of detecting TeV blazars would be enhanced, since 
its effective 
beaming angle is actually more like the jet opening 
angle.  
Also, the wider jet would mean a bigger reduction in the apparent velocity of the 
radio knot.  A key implication is that the combination of these
factors is probably responsible for the surprising preference of TeV blazars
for possessing VLBI knots displaying slower motions.

\acknowledgments
PJW thanks the organizers for arranging a superb meeting;
his efforts were
supported in part by NSF grant AST-0507529.

                                                                                   \label{lastpage}
\end{document}